\documentclass[twocolumn,prl,showpacs]{revtex4}

\newcommand{\be}{\begin{equation}}
\newcommand{\ee}{\end{equation}}

\begin {document}

\title{Vortex mass in a superfluid at low frequencies}
\author{D. J. Thouless}  
\affiliation{Department of Physics, Box 351560, University of Washington,
Seattle, Washington 98195}
\author{J. R. Anglin}
 \affiliation{Fachbereich Physik, Technische Universit\"at Kaiserslautern, 67663 Kaiserslautern,
 Germany}
 
\date{\today}

 \begin{abstract} An inertial mass of a vortex can be calculated by 
driving it round in a circle with a steadily revolving pinning potential.  
We show that in the low frequency limit this gives precisely the same 
formula that was used by Baym and Chandler, but find that the result is not unique and depends on the force field used to cause the acceleration.  We apply this method to the Gross-Pitaevskii model, and derive a simple formula for the vortex mass.  We study both the long range and short range properties of the solution.   We agree with earlier results that the non-zero compressibility  leads to a divergent mass.  From the short-range behavior of the solution we find that the mass is sensitive to the form of the pinning 
potential, and diverges logarithmically when the radius of this potential 
tends to zero. \end{abstract}
  
 \pacs{PACS: 67.40.Vs, 47.37.+q }

\setcounter{equation}{0}
 
\maketitle
  \section{Introduction}

  Conflicting results on the mass of a quantized vortex in a neutral 
superfluid can be found in the literature.  Popov \cite{popov} and Duan 
\cite{duan94} argue that the mass per unit length is infinite, while Baym 
and Chandler \cite{baymchandler} argue that it is negligible.  We have 
developed a method for studying the problem of vortex mass and applied it 
to the Gross-Pitaevskii model \cite{gross,pitaevskii} for superfluids, 
which is known to describe weakly interacting bosonic atoms well, and is 
believed to give a useful qualitative description of superfluid $^4$He.  
We have obtained new analytical results and laid the basis for detailed 
numerical calculations.  Firstly, we show that our method is equivalent 
to the method described by Baym and Chandler.  Secondly, we specialize to 
the Gross-Pitaevskii model and derive a new and compact formula for the 
vortex mass.  Thirdly, we agree with Popov and Duan that there is a divergent contribution to the mass due to the expansion of the fluid at large distances from the vortex core.  Fourthly, we get the new result that the mass is sensitive to the form of the pinning force that sustains the motion of the vortex in the presence of the Magnus force, and that the mass diverges as the range of this pinning force tends to zero.  This suggests that an unambiguous vortex mass may not exist, and that inertial effects in vortex dynamics may be scenario-dependent \cite{anglin07}.

Vortex motion is important in many physical scenarios, from collective modes in neutron star matter\ref{bc} to decay of superconducting currents via quantum nucleation of vortices\cite{vinen61} and quantum turbulence in superfluids\cite{vd}.  The universal topological constraints on vortex behavior are an important asset in understanding these diverse and important phenomena; vortex inertial mass is a salient uncertain parameter in the otherwise highly constrained vortex dynamics.  Since theories of a universal vortex mass based on different assumptions do not agree, our contribution here is to consider a specific scenario in which a vortex mass may be identified without ambiguity.

  To ascertain the mass of an object in the conventional way one 
determines the acceleration due to an applied force.  Most of the methods that have been used so far to determine the mass of a vortex in a superfluid have, however, avoided the complexity of actual vortex acceleration.  In Popov's \cite{popov} work analogies between superfluidity and electrodynamics are used, where vortices correspond  to lines of electric charge and sound waves to electromagnetic waves.  The energy $E$ 
needed to form a vortex at rest is calculated, and the mass $M_V$ is found by using the Einstein formula $E=M_Vc_s^2$, where $c_s$ is the sound velocity in the superfluid.  The long range of the field due to a line charge leads to an infrared divergence of the self-energy, and so to an infrared divergence of the vortex mass.  Duan gets similar results which he argues are due to the breaking of gauge invariance for the neutral superfluid.   In the Baym and Chandler work \cite{baymchandler} the calculation of mass is based on a calculation of the extra energy associated with a vortex forced to move with a constant velocity relative to the fluid.  The authors argued that this energy is primarily associated with the core of the vortex, and that this is small, but they did not describe a detailed calculation.

Accelerating vortices can be analyzed in some theories, however.  In classical incompressible hydrodynamics the mass of a hard-cored vortex can be calculated from the resonant frequency at which the vortex can move, without any applied force, on a circular orbit\cite{lamb}.  The mass obtained is just the finite mass of fluid displaced by the vortex core, plus the mass of the material contained in the core.  In the Gross-Pitaevskii model, it has been shown that damping due to radiation of phonons is too strong to leave any trace of a similar resonance, unless the vortex core is loaded with a large additional mass \cite{dan96,quist}.  We have carried out a more general and detailed calculation similar to that of Quist \cite{quist}, and we get similar results, even when the bulk modulus is much higher than it is in the Gross-Pitaevskii model \cite{anglin07}.  

In the classical incompressible fluid, the same vortex mass can also be found by analyzing circular vortex motion under an applied external force (acting on the core).  A vortex moving with constant velocity $v$ relative to the fluid must be acted on by a transverse force, equal to 
$\rho_s\kappa v$ per unit length, where $\kappa$ is the circulation round 
the vortex and $\rho_s$ the superfluid density.  This is the Magnus effect, which follows from conservation of momentum and circulation, and so applies to vortices in quantum fluids as well as classical.  We believe it may be an important omission in earlier work, that remarks such as ``Now give the vortex an instantaneous velocity $\bf v$'' \cite{duan94,baymchandler} are made, without considering how the necessary force is to be applied.  We therefore consider an external  `pinning' potential, which repels the fluid particles, and hence attracts the low-density vortex core.  By moving this potential around a circular orbit, we drag the vortex.  We take it as our definition of the vortex mass,  that the force required to do so will differ from the Magnus force by the centripetal force, equal to the product of vortex mass and acceleration.  An orbiting vortex will generally excite waves, requiring an azimuthal force to maintain constant speed against radiative damping; we avoid this issue by considering slow enough orbits that this effect becomes negligible. 

We therefore begin by considering the perturbatively slow circular motion of a
driven, pinned vortex in a homogeneous quantum fluid, and by showing that our 
mass definition gives the Baym-Chandler formula \cite{baymchandler} in the low 
orbit frequency limit.  

\section{Perturbation theory in a rotating frame}
We consider a many-body hamiltonian $H$ which 
includes the kinetic energy of the atoms, the translation-invariant 
interaction $U_I$ between them, and a moving pinning potential $V_p({\bf r},t)$, 
which we take to have an instantaneous axis of symmetry.  We take it to move 
round a circle of radius $a$ at a constant angular frequency $\omega$.  
We ignore the confining boundaries that hold the system in place and hold the density constant.  In a coordinate system that rotates with the pinning potential the Hamiltonian becomes time-independent;  using polar co-ordinates centered on the axis of the 
pinning potential, rather than on the center of the circular orbit, we replace $\partial_t$ by $ \partial_t+v\partial_y+\omega\partial_\phi$,
where $v=a\omega$.   We therefore consider an eigenstate of the 
rotating, shifted frame hamiltonian,
  \be\label{eq:hamilton} H_R=H+\omega(i\partial_\phi+1)+iv\partial_y \;,\ee
  where we have chosen units with $\hbar=1$.  We write $H'=H_R-iv\partial_y$ and  treat the term proportional to $v$ as a perturbation. 
The vortex state for $v=0$ is $|\Psi_0\rangle $, with total energy $E_0$. 
To leading order in $v$ we thus have
  \begin{eqnarray}|\Psi_{v}\rangle &=& \left[1+( E_0-H')^{-1}(iv\partial_y)\right] |\Psi_0\rangle\\
  E_{v}&=&E_{0}+v^{2}\langle\Psi_{0}|\partial_y( E_0-H')^{-1}\partial_y|\Psi_0\rangle \;.\end{eqnarray}
The expectation value of the force exerted by the pinning potential is, 
to lowest order in $v$,
  \be\label{eq:force11} F_x= 2v\Im \langle\Psi_0|(\partial_xV_p) 
(E_0-H')^{-1}\partial_y |\Psi_0\rangle \;.\ee
  We can use the commutation relation
  \be\label{eq:commute}[\partial_x,H']= \partial_xV_p +i\omega \partial_y 
\ee
  to rewrite this as
  \begin{eqnarray}\label{eq:force12} F_x&=& iv\langle\partial_x \psi_0| \partial_y\psi_0\rangle 
-iv\langle\partial_y \psi_0|\partial_x \psi_0\rangle
\nonumber\\&&+2v\omega 
\langle\psi_0|\partial_y( E_0-H')^{-1}\partial_y|\psi_0\rangle \;. \end{eqnarray}
The first two terms on the right can be shown, by applying Stokes 
theorem, to give $\rho v$ times the circulation \cite{tan96}, and this is 
the standard form for the Magnus force. The coefficient of the acceleration $v\omega$ is our
vortex mass $M_{V}$.  Comparing it with (3) above, we recover the formula of Baym and Chandler \cite{baymchandler}, 
\begin{equation}
M_{V}=-\frac{\partial^{2}E_{v}}{\partial v^{2}}\;,
\end{equation}
but now derived from the force on an accelerating vortex.  
  
Rather than proceeding further with the full many-body problem, we specialize to the Gross-Pitaevskii mean field theory \cite{gross,pitaevskii}.  We study the asymptotic properties of the solution for large distances, and display the divergent contribution to the vortex mass for this compressible quantum fluid.  We then study the 
short-range properties and consider the influence of the form of the 
pinning potential on the vortex mass.  We find that the 
vortex mass depends strongly on the pinning potential, and diverges when 
its radius goes to zero.  Finally, we briefly
consider the implications of these results.

\section{Gross-Pitaevskii equation and compressibility}

We consider the 
Gross-Pitaevskii equation in a frame of reference moving with speed $v$, 
which takes the form, in appropriately rescaled units\cite{unitsnote},
  \be\label{eq:gprotate} -\nabla^2\psi +V_p(r)\psi +(|\psi|^2-1)\psi 
=-iv\partial_y\psi\;. \ee

The vortex solution for the limit $v\to 0$ has the form studied by Gross and Pitaevskii 
\cite{gross,pitaevskii}, $\psi_0(r) \exp(i\phi)$, where $\psi_0$ satisfies
  \be\label{eq:vortex} -\psi_0''-{1\over r}\psi_0'+{1\over 
r^2}\psi_0+(V_p+|\psi_0|^2-1)\psi_0 =0\;; \ee
  we take $\psi_0$ to be real.  We use the methods of Fetter \cite{fetter}, so that,  to first order in perturbation theory, we write the solution as
 \be\label{eq:wavefunction} \psi_{v} \approx \psi_0(r) e^{i\phi} +\sum_{m=0,2}
v \chi_m(r)e^{im\phi} \;.\ee
  We can substitute this back into
Eq.\ (\ref{eq:gprotate}) to get the equation, to first order in $v$,
 \begin{eqnarray}  \label{eq:inhomo0}\sum_{m=0,2} [-\chi_m''-{1\over r}\chi_m'+ {m^2\over
r^2}\chi_m \!\!&+&\!\!(V_p+2\psi_0^2-1+ i^m\omega)\chi_m\nonumber\\
 +\psi_0^2\chi^*_{2-m}]  e^{im\phi}
  =&&\!\!\!\!\!\!\!\!{\cos\phi\over r} \psi_0 -i\sin\phi\;\psi_0' \;;\end{eqnarray}
 we can take $\chi_m$ to be real.  In this paper we discuss the limit $\omega\to 0$ of Eq.\ (\ref{eq:force12}), and in this limit the Galilean invariance of Eq.\ (\ref{eq:vortex}) in the region where the pinning potential is zero gives a particular integral 
 \be\label{eq:pi1} \chi_0=-{1\over 4}r\psi_0\,,\ \chi_2={1\over 
4}r\psi_0\;. \ee
 This satisfies the boundary conditions for small $r$, but a solution 
$(f_0,f_2)$ of the homogeneous equation
  \be\label{eq:homo0} -f_m''-{1\over r}f_m'+ {m^2\over r^2}f_m 
+(V_p+2\psi_0^2-1)f_m +\psi_0^2f_{2-m}=0 \ee
must be added to meet the boundary condition at $r\to\infty$.
 
  For large $r$ we use  $f_\pm=f_0\pm 
f_2$, which satisfy the homogeneous equation
   \be\label{eq:gp2} \left(\begin{array}{cc} H_{++}&H_{+-}\cr 
H_{-+}&H_{--}\end{array} \right)
  \left(\begin{array}{c} f_+\cr f_- \end{array} \right) = 0 \;,\ee
  where\begin{eqnarray}\label{eq:gp3}
  H_{+-} &=&H_{-+} =-{2\over r^2}\;,\nonumber\\
 H_{++}&=&-{d^2\over dr^2}-{1\over r}{d\over dr} +{2\over 
r^2} 
+V_p+3\psi_0^2 -1\,\;,\\
 H_{--}&=&-{d^2\over dr^2}-{1\over r}{d\over dr} +{2\over 
r^2} +V_p+\psi_0^2 -1 \;.\nonumber\end{eqnarray}
    The two solutions of the homogeneous equation bounded at large $r$ fall 
off like $f_-\sim 1/r$ and $f_+\sim \exp(-\sqrt 2 r)/\sqrt r$, and there is an unbounded 
solution with $f_-\sim r$, which can be used to cancel the dominant term 
in $-r\psi_0/2$.  We thus look for a bounded solution of Eq.\ (\ref{eq:inhomo0}) of the form
  \be \chi_+=f_+\,,\ \ \chi_-=f_--r\psi_0/2\;.\ee 
  
  To use the Baym-Chandler formula for the centrifugal force in Eq.\ 
(\ref{eq:force12}), we can calculate the perturbed wave function to first 
order in $v$, and then calculate the excess energy that it contributes.   
This gives,  with the use of Eqs.\  (\ref{eq:vortex}), (\ref{eq:gp2}), 
(\ref{eq:gp3}),
  \begin{eqnarray}\label{eq:mass}
M_V&=& 4\pi\int_0^\infty \left(f_+\ \  f_--{1\over 2}r\psi_0\right) \nonumber\\
&&\times   \left(\begin{array}{cc} H_{++}&H_{+-}\nonumber\\
H_{-+}&H_{--}\end{array} \right)
  \left(\begin{array}{c} f_+\cr f_- -{1\over 2}r\psi_0 \end{array} \right)  
r\,dr \;,\nonumber\\
&=&4\pi\int_0^\infty [(f_--{1\over 2}r\psi_0 ) 
\psi_0'+{f_+\over r}\psi_0] r\,dr   \;.\end{eqnarray}
  
 To find out whether the effects  of the compressibility at large $r$ give 
rise to a divergent vortex mass, as Popov \cite{popov} and Duan 
\cite{duan94} have argued, we must find the asymptotic behavior of 
solutions of Eq.\ (\ref{eq:gp2}) for large $r$ and substitute this into 
Eq.~(\ref{eq:mass}).  We have $\psi_0\sim 1-1/2r^2$.  Bounded solutions of the homogeneous equation have the form
 \be f_+\sim {1\over 2r}\,,\ \ f_--{1\over 2}r\psi_0\sim {1\over 2r}\ln r 
+{a_1\over 2r}\;;\ee
  the coefficient $a_1$ is determined by the small $r$ boundary.  The potential energy given by the $f_+$ term gives an integrand proportional to $r^{-1}$, and so gives the logarithmically divergent expression for mass found by Popov \cite{popov} and Duan \cite{duan94}.  For real systems this integral will be cut off at large distances by the finite size of the system, the presence of other vortices, or by non-zero frequency effects.
  
  \section{Influence of the pinning potential}

  To study the influence of the pinning potential $V_p$ we take the 
specific case of a hard core repulsive potential of radius $r_c$, so that 
$\psi_0(r)=0$ for $r\le r_c$.  The particular integral given by Eq.\ 
(\ref{eq:pi1}) for the $\omega\to 0$ limit vanishes at $r=r_c$, so we must 
add solutions of the homogeneous equation that also vanish there in order 
to satisfy the boundary conditions for large $r$.  To study these we go 
back to the representation used in Eqs.\ (\ref{eq:inhomo0}), 
(\ref{eq:pi1}), (\ref{eq:homo0}). 
 If $r_c<<1$, two independent solutions 
for small $r$ that vanish at $r=r_c$ can be constructed by combining 
solutions regular and irregular at the origin, and, close to $r_c$ where  
$\psi_0^2$ is small, these have the forms
  \begin{eqnarray}\label{eq:smallr} 
f_0^{(1)} &\approx& -(\ln{r_c\over 2}+\gamma)J_0(r) -{\pi\over 2}Y_0(r)\,,\ \  f_2^{(1)}\approx 0\;,\nonumber\\
 f_0^{(2)} &\approx& 0\,,\ \ f_2^{(2)}  \approx {2\over 
r_c^2}J_2(r) +\pi{r_c^2\over 16} Y_2(r) \;;\end{eqnarray}
 Here $J,Y$ are Bessel functions and $\gamma$ is the Euler constant.  The 
solution to lowest order in $\psi_0$ is then
\begin{eqnarray} \label{eq:soln}
\psi\approx e^{i\phi}\{\psi_0 \!+\! v[{ir\over 
2}\psi_0\sin\phi\! 
 +\!a_1 \chi_0^{(1)}e^{-i\phi} \!+\!a_2 \chi_2^{(2)}e^{i\phi}] 
\},\ \ \end{eqnarray}
  where $a_1,a_2$ are real.
  
   Determination of these two coefficients requires a detailed integration 
of the homogeneous part of Eqs.\ (\ref{eq:gp2}) and (\ref{eq:gp3}), but 
there is an important constraint placed on them by the fact that the 
Magnus force has magnitude $2\pi v$.  For a hard core potential of radius 
$r_c$ the force per unit length can be written as
   \be\label{eq:hardforce} F_x 
=r_c\int_0^{2\pi}|\partial_r\psi(r_c,\phi)|^2 
\cos\phi \,d\phi \;.\ee
   From Eqs.\ (\ref{eq:soln}) and (\ref{eq:smallr}) we get, at $r=r_c$,
    \be\label{eq:force3}\Re \partial_r (\psi e^{-i\phi})= \psi_0'+{v\over 
r_c}(a_1 +a_2)\cos\phi\;, \ee
    and the known value of the Magnus force gives
     \be\label{eq:magnus} \psi_0'(r_c)(a_1  +a_2)=1\;. \ee
     
     Energy considerations show that $a_2$ should be small, since a value 
of order unity would make a contribution of order $r_c^{-2}$ to Eq.\ 
(\ref{eq:mass}), so $a_1$ must be close to $\pm 1$.  The dominant 
contribution to  Eq.\ (\ref{eq:mass}) is
      \be\label{eq:mass2} M_V \approx 4\pi\int_{r_c}^R a_1\ln r_c 
J_0(r)(\psi_0'+{\psi_0\over r})dr \;,\ee
      where $R$ is a length of the order of unity. This shows that the 
vortex mass is sensitive to the form of the pinning potential, and diverges 
logarithmically as the core radius of the pinning potential tends to zero.  It is the high quantum pressure near $r_c$ that leads to the large kinetic energy for small $r_c$.
     
    We were actually led to this result by a different approach in which 
we ignored the nonlinearity of the equation for small $r$, and matched the 
solution to a region of incompressible flow for large $r$.  This 
simplified model allowed us to get an exact expression of the vortex mass 
which displays this logarithmic divergence, and for which the wave 
function has the same general features at small $r$.  This will be discussed elsewhere \cite{anglin07}.

\section{Discussion}
   
  We have shown how to obtain an expression for the inertial mass of a 
stable quantized vortex in an infinite neutral superfluid by subjecting it 
to a straight, circularly symmetric, pinning potential $V_p$ which is 
slowly and steadily rotated about a parallel axis whose distance $a$ from 
the vortex is large compared with the size of the vortex core. The vortex 
has a steady state in a frame of reference that rotates about this axis of 
rotation, with the same angular velocity $\omega$ that the pinning 
potential rotates.  We use perturbation theory to study this steady state, 
and to find the force which the pinning potential exerts on the vortex to 
keep it in a steady rotation. The leading term in this expansion, 
proportional to $v$, gives the Magnus force, in a form which is closely 
analogous to the Magnus force acting on a moving vortex in classical fluid 
mechanics.  The next term, proportional to $v\omega$, has a coefficient 
that can be interpreted as an inertial mass of the vortex.

  It is essential to have a pinning potential to stabilize the position of 
the vortex in a rotating system, and it has to be strong enough to hold the 
vortex against the Magnus effect and the centrifugal force.  We agree with Popov and Duan that the mass determined this way is logarithmically divergent in the low-frequency limit, and we have shown that it depends sensitively on the form of the pinning potential, diverging logarithmically as the radius of the pinning potential tends to zero. 

 An important tentative conclusion of this work is that ``the mass'' of a vortex is not well defined, but depends on the process by which the mass is measured.  We shall discuss this further in more detail elsewhere.
   
  \begin{acknowledgments}   We are grateful for the hospitality of the Aspen Center for Physics, where this work was started, and for the partial support provided by the U.S. National Science Foundation, award DMR-0201948. \end{acknowledgments}


\begin{thebibliography}{99}
\bibitem{unitsnote} When we calculate the vortex mass per unit length in these units, the result has to be multiplied by $\hbar^{2}/g$, where $g$ is co-efficient of the $|\psi|^{2}\psi$ term in the unrescaled equation.  For a dilute Bose gas, $\hbar^{2}/g = m/(4\pi a)$, where $m$ is the particle mass and $a$ the s-wave scattering length.  In general $\hbar^{2}/g = \rho_{0}\xi^{2}$, where $\rho_{0}$ is the ambient mass density and $\xi$ the so-called healing length.
\end{thebibliography}
\end{document}